\documentclass{Interspeech2024}
\usepackage{csquotes}
\usepackage{multirow}
\usepackage{adjustbox}
\usepackage[inkscapelatex=false]{svg}
\usepackage{caption}
\usepackage{subcaption}
\usepackage{stackengine}
\usepackage{amsmath}
\usepackage{hyperref}

\interspeechcameraready

\title{Hear Your Face: Face-based voice conversion with F0 estimation}

\name[affiliation={1}]{Jaejun}{Lee}
\name[affiliation={1}]{Yoori}{Oh}
\name[affiliation={1}]{Injune}{Hwang}
\name[affiliation={1,2,3}]{Kyogu}{Lee}

\address{
  $^1$Department of Intelligence and Information, Seoul National University\\
  $^2$Interdisciplinary Program in Artificial Intelligence, Seoul National University \\
  $^3$Artificial Intelligence Institute, Seoul National University}
\email{\{jjlee0721, yoori0203, dlswns8, kglee\}@snu.ac.kr}


\newcommand{\etal}{\textit{et al.}}

\begin{document}

\maketitle

\begin{abstract}
This paper delves into the emerging field of face-based voice conversion, leveraging the unique relationship between an individual's facial features and their vocal characteristics.
We present a novel face-based voice conversion framework that particularly utilizes the average fundamental frequency of the target speaker, derived solely from their facial images.
Through extensive analysis, our framework demonstrates superior speech generation quality and the ability to align facial features with voice characteristics, including tracking of the target speaker's fundamental frequency.
\end{abstract}

\section{Introduction}
Voice, the cornerstone of human speech, plays a crucial role in interpersonal communication. Beyond its communicative function, voice is a distinctive feature of an individual, reflecting personal identity.
Consequently, individuals who are unable to produce sound not only face a significant barrier to communication but also experience a loss of personal expression.

Conventional speech synthesis techniques, such as Text-to-Speech~(TTS) and Voice Conversion~(VC), have made significant strides in emulating a target voice while retaining the non-verbal content elements. Yet, these techniques predominantly rely on the availability of the target voice's acoustic data to replicate its unique speech style effectively.

The human face represents another intrinsic aspect of individual identity, containing details such as biological gender, ethnicity, and age. More than just exploring visual information from face, recent studies have increasingly focused on understanding the relationship between facial features and vocal attributes \cite{kim2019learning, oh2019speech2face}. This field of study may hold the key to a new form of speech synthesis, one that retains the target speaker's identity even in the absence of vocal information.

Recent advancements in face-based speech synthesis have experienced a notable surge, particularly through the integration of conventional TTS \cite{goto2020face2speech, pluster2021hearing, lee2023imaginary, wen2021seeking} and VC \cite{lu2021face, takahashi2023cross, sheng2023face} techniques. While this growing interest and development show a promising view, the field remains in its formative phases. Specifically, identifying a `voice that matches the face' presents significant challenges, and metrics for evaluating it also remain a pivotal question.

The fundamental frequency ($F0$), one of the key components in voice conversion process \cite{choi2021neural, choi2022nansy++}, not only serves as a pitch information of speech but also has an aspect of containing information of speakers identification \cite{xu2013acoustic}.
It has been found that facial features have correlation with voice pitch information, even in cases where gender is controlled \cite{kim2019learning}.
It implies that voice pitch information, indicated by the $F0$, could be derived from the speaker's facial images, and it is not merely from basic gender identification, but also from further biological associative information.

In this study, we propose a novel framework for face-based speech synthesis, focusing particularly on the voice conversion that imprints a face-based target voice's characteristics onto the original source audio. Our framework specifically utilize the $F0$ of target speaker, derived solely from facial images. This approach aims to enhance the face-based voice conversion process, generating speech that is well aligned with the target individual's vocal identity without using any acoustic data of the target speaker.

To contextualize our research, we delineate our contributions as follows:
\begin{itemize}
\item We present a framework that sets a new benchmark in performance for face-based voice conversion, demonstrating state-of-the-art results.
\item We propose a novel approach for speech synthesis by estimating the $F0$ of the target speaker through their facial images.
\item Through extensive analysis and the introduction of a novel evaluation metric, we demonstrate that our framework not only produces high-quality synthetic speech but also suggests that the synthesized voice aligns reasonably well with the corresponding facial image.
\end{itemize}

The demo is available on the link, \url{https://jaejunL.github.io/HYFace_Demo/}.

\section{Related work}

\subsection{Voice conversion}
Voice conversion, a specialized subset of speech synthesis, is a process that automatically transforms speech from one source speaker into a voice resembling that of a target speaker, all the while maintaining the original linguistic content.
The challenge primarily arises in non-parallel voice conversion scenarios, where the lack of directly corresponding parallel data.
The disentanglement of linguistic content in speech and its acoustic voice, timbre is a crucial problem. To address this, various methodologies have been explored, including adversarial training \cite{hsu2016voice, qian2019autovc}, vector quantization \cite{liu2019unsupervised, wang2021vqmivc}, and information perturbation \cite{choi2021neural, choi2022nansy++}.

\begin{figure*}[t!]
\centerline{\includegraphics[width=2.12\columnwidth]{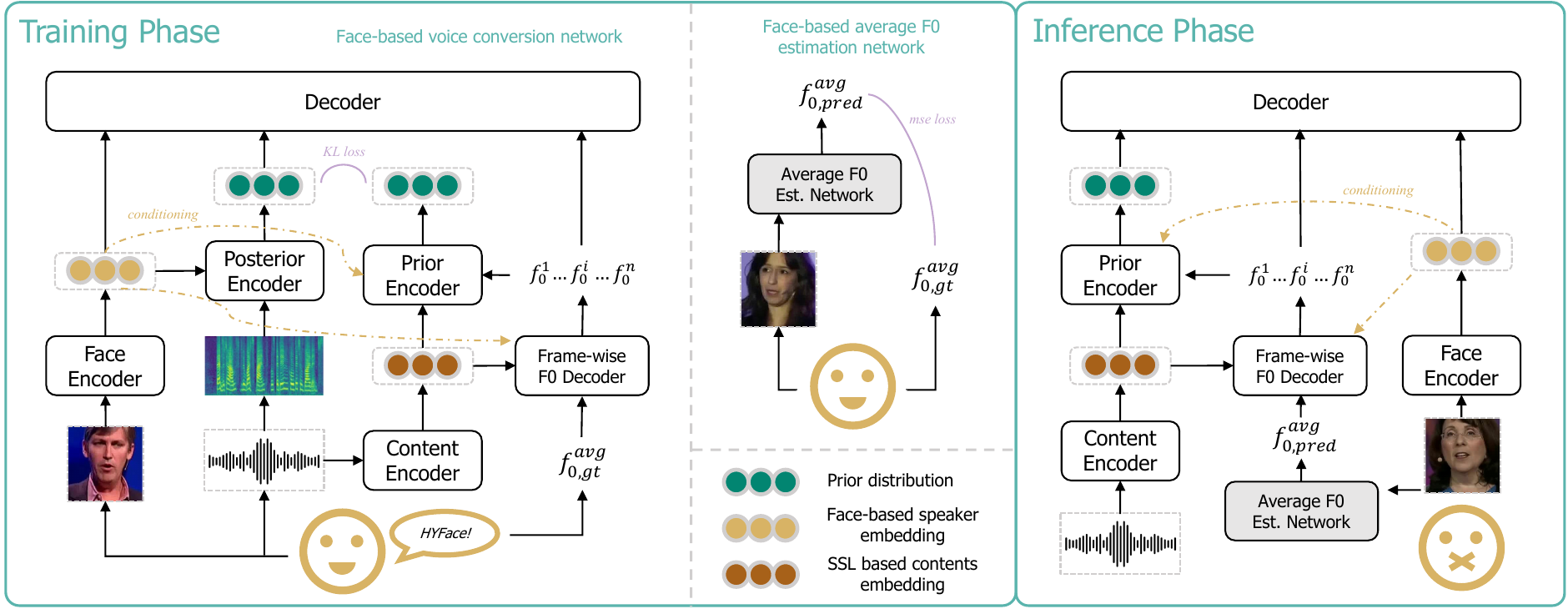}}
\centerline{\hspace{2.3cm}(a) Training phase\hspace{6cm} (b) Inference phase}
\caption{Overview of the proposed method, HYFace, conditional VAE based network that its speaker embedding is learned on face images only. In training phase, a predefiend speaker-wise average F0 ($f_{0,\mathit{gt}}^{\mathit{avg}}$) is used to estimate frame-wise $F0$ values. However, as the $f_{0,\mathit{gt}}^{\mathit{avg}}$ values for unseen target speakers are not available during the inference phase, we independently train an average $F0$ estimation network based solely on facial inputs. This module is then utilized in the inference phase.}
\label{fig:figure1}
\end{figure*}

Notably, recent advancements have been made with the advent of self-supervised learning (SSL) techniques. Pretrained representations trained on large data corpus exhibit a remarkable capacity for disentangling the contents information of speech \cite{hsu2021hubert, qian2022contentvec}, thereby significantly enhancing voice conversion processes \cite{van2022comparison}. Recently, the collaborative voice conversion project, named Sovits-SVC\footnote{https://github.com/svc-develop-team/so-vits-svc} (SoftVC VITS Singing Voice Conversion), has demonstrated impressive outcomes in both standard voice conversion and singing voice conversion domains. It leverages SSL representations for content representations, and employs a neural-source filter vocoder, specifically designed to track the $F0$ of the source audio, which plays a significant role in its original intention for singing voice conversion.

\subsection{Face-voice association}
Early studies, especially through human behavioral and neuroimaging approach, demonstrate that humans use both facial and vocal cues for identity recognition \cite{xu2013acoustic, campanella2007integrating}. Furthermore, similar studies reveal human ability to match faces with voices of unfamiliar individuals \cite{kim2019learning, mavica2013matching, smith2016matching}. Particularly, the authors in \cite{kim2019learning} showed that humans can significantly match faces and voices under the controlled attributes such as gender, race, and age. Specifically, they also revealed that there is a correlation between the target speaker's voice pitch and facial features.

Building on these finding, interest has surged in learning based methods for associations between faces and voices. An application of such methods includes generating face from a given speech \cite{oh2019speech2face, choi2019inference} or vice versa. Specifically, face-based speech synthesis, the focus of this paper, is categorized based on the type of input: text for TTS \cite{goto2020face2speech, pluster2021hearing, lee2023imaginary, wen2021seeking} and source audio for VC \cite{lu2021face, takahashi2023cross, sheng2023face, wang2022residual}. Recently, Sheng \etal~\cite{sheng2023face} showed prominent result in zero-shot face-based voice conversion, employing memory based methods. 
All these works tried to learn cross-modal speaker representations implicitly, without explicit voice characteristic such as $F0$. Moreover, their evaluation primarily relied on metrics such as the mean opinion score (MOS) or speaker embedding similarities, rather than on assessments directly related to explicit voice characteristics.

\begin{table*}[t!]
\caption{Evaluation result. The definitions of all metrics are provided in Section \ref{experiment3}.}
\centering
\begin{adjustbox}{width=0.95\textwidth}
\begin{tabular}{ccccccccccccc}
\hline\hline
\multirow{2}{*}{} & \multicolumn{2}{c}{Homogeneity$\uparrow$} & \multicolumn{2}{c}{Diversity$\downarrow$} &
\multicolumn{2}{c}{Consistency(obj)$\uparrow$} & \multicolumn{2}{c}{Consistency(sub)$\uparrow$} &
\multicolumn{2}{c}{Naturalness$\uparrow$} & \multicolumn{2}{c}{ABX test($\%$)$\uparrow$}\\
\cmidrule(lr){2-3} \cmidrule(lr){4-5} \cmidrule(lr){6-7} \cmidrule(lr){8-9} \cmidrule(lr){10-11} \cmidrule(lr){12-13}
 & HMG & HTG & HMG & HTG & HMG & HTG & HMG & HTG & HMG & HTG & HMG & HTG \\
\hline
GT &  0.7456 & - & 0.5418 & - & - & - & 3.9048 & - & 4.0469 & - & - & -    \\ \hline
FVMVC\cite{sheng2023face} &  0.6391 & 0.6401 & \textbf{0.5942} & \textbf{0.5976} & 0.5105 & 0.5086 & 3.5705 & 3.5009 & 3.4096 & 3.2470 & 0.395 & 0.420 \\ 
HYFace & \textbf{0.6770} & \textbf{0.6793} & 0.6072 & 0.6103 & \textbf{0.5696} & \textbf{0.5632} & \textbf{3.8916} & \textbf{3.8189} & \textbf{3.8313} & \textbf{3.7651} & \textbf{0.605} & \textbf{0.580} \\ 

\hline
\hline
\end{tabular}
\end{adjustbox}
\label{table:result}
\end{table*}

\section{Methods} \label{sec:method}
In this section, we present our proposed method, HYFace (short for `Hear Your Face'), a novel approach to face-based voice conversion, it begins in Section \ref{subsec:method1}.
Then, Section \ref{subsec:method3} provides detailed architectures of our proposed model.
Figures \ref{fig:figure1}(a) and \ref{fig:figure1}(b) illustrate the procedures of the training phase and the inference phase of our method, respectively.

\subsection{HYFace} \label{subsec:method1}
Our HYFace network is a voice conversion (VC) framework fundamentally inspired by Sovits-SVC, utilizing a conditional variational autoencoder architecture. It incorporates pretrained SSL representations as content input for the prior encoder.
However, distinct from traditional VC frameworks, HYFace uses the facial image of the target speaker to modify the style of the source audio, instead of using the target speaker's voice. In this system, the speaker embedding, which is learned from the facial images, conditions the prior encoder, posterior encoder, decoder and the frame-wise $F0$ decoder ($\mathit{FF}$).

Additionally, to enhance the model's capacity to incorporate target voice characteristics, frame-wise $F0$ values, $f_0^i$ ($i=1,...,n$. $n$ is the number of frames) conditions both the prior encoder and the decoder.
A speaker-wise average $F0$~($f_{0,\mathit{gt}}^{\mathit{avg}}$) 
is adjusted to $f_0^i$ within the $\mathit{FF}$, in conjunction with the content embedding ($c$) and face-based speaker embedding ($s$).
The loss~$\mathcal{L}_{\mathit{ff}}$ for training $\mathit{FF}$ is as follows:
\vspace{-2mm}
\begin{equation}
\mathcal{L}_{\mathit{ff}}=\frac{1}{n}\sum_{i=1}^n(f_{0,\mathit{gt}}^i-\mathit{FF}(f_{0,\mathit{gt}}^{\mathit{avg}}, c, s))^2,
\vspace{-2mm}
\end{equation}
where $f_{0,\mathit{gt}}^i$ refers to the ground-truth frame-wise $F0$ values.
Note that $f_{0,\mathit{gt}}^{\mathit{avg}}$ value represents the average of $f_{0,\mathit{gt}}^i$ values across all audio frames for each speaker in the training dataset.
Importantly, due to the unavailability of $f_{0,\mathit{gt}}^{\mathit{avg}}$ information for unseen target speakers during the inference phase, we independently train a face-based average $F0$ estimation network ($\mathit{AF}$) solely on face image ($v$) of target speakers, which constitutes one of the key components of our proposed method. The loss~$\mathcal{L}_{\mathit{af}}$ for training $AF$ is as follows:
\begin{equation}
\mathcal{L}_{\mathit{af}}=(f_{0,\mathit{gt}}^{\mathit{avg}}-\mathit{AF}(v))^2.
\end{equation}

Then, $\mathit{AF}$ is utilized during the inference phase, enhancing our face-based voice conversion network to produce speech that better aligns with the voice characteristics of the target speaker.
To clarify our HYFace training procedure, we describe our other loss functions, which include reconstruction loss, KL~(Kullback-Leibler) divergence loss, adversarial loss, and feature matching loss in Supplementary A.
              
\subsection{Model architecture} \label{subsec:method3}
This section details the architecture of the modules employed in our model. We note that all modules were trained from scratch, except for the contents encoder, which is based on pretrained models.

\noindent\textbf{Posterior Encoder}: It is consists of WaveNet-based residual blocks. To integrate face-based speaker embeddings, we employed global conditioning similar to \cite{kim2021conditional}.\\
\textbf{Prior Encoder}: It has a transformer-based architecture similar to \cite{shaw2018self}, atop which is stacked a normalizing flow layer comprised of residual coupling blocks \cite{dinh2016density}.\\
\textbf{Face Encoder}: Vision Transformer \cite{dosovitskiy2020image} architectures with projection layer.\\
\textbf{Contents Encoder}: We used ContentVec \cite{qian2022contentvec}, pretrained SSL represenations, especially hugging face version\footnote{https://huggingface.co/lengyue233/content-vec-best}\\
\textbf{Decoder}: It basically has architecture similar to the generator of HiFi-GAN \cite{kong2020hifi} so as to our discriminator network ($D$), but for careful conditioning of $F0$ information, we used a neural source filter method \cite{wang2019neural} based conditioning similar to Sovits-SVC.\\
\textbf{Frame-wise $\textbf{F0}$ Decoder ($\textbf{FF}$}): It is based on architecture with self-attention layers and feed forward layers conditioned with both content embedding and face-based speaker embedding. Fast Context-base Pitch Estimator\footnote{https://github.com/CNChTu/FCPE} (FCPE) is used to extract frame-wise $F0$ value and speaker-wise average $F0$ value.\\
\textbf{Average $\textbf{F0}$ estimation network (AF)}: Similar to face encoder, vision transformer based architectures with projection layer.\\

\section{Experiments}
\subsection{Dataset} \label{experiment1}
We used LRS3 \cite{afouras2018lrs3}, the dataset consists of 5,502 videos from TED and TEDx which has more than 430 hours long. Each video is cropped on speaker's face and it has a resolution of 224×224 with 25 frames per seconds images and 16kHz single channel audio. We used predefined \textit{pretrain}, \textit{trainval} and \textit{test} set for training, validation and evaluation, respectively. Expecting our proposed model to more carefully associate detailed face features with speaker's voice characteristics, we used only frontal images from the dataset. Especially, we employed OpenCV haarcascades\footnote{https://github.com/opencv/opencv/tree/master/data/haarcascades} for image selection, resulting in about 20\% of image data being filtered out.

For the evaluation, we picked 50 male and 50 female speakers on \textit{test} set, ranked by the amount of data available. On average, there are about 5.8 speech audio files and 280 facial images available per speaker.
We hypothesized that converting the voice to a target speaker of a different gender from the source is more challenging than converting to the same gender. 
Therefore, we constructed two types of evaluation sets: Homogeneous Gender (HMG) set and Heterogeneous Gender (HTG) set. The HMG set pertains to face-based voice conversion scenarios in which the target speaker's gender is the same as the source speaker's (either male to male (M2M) or female to female (F2F)). In contrast, the HTG set applies to scenarios where the target speaker's gender is different from that of the source speaker (from male to female (M2F) or female to male (F2M)). Thus, technically we have four evaluation sets: M2M and F2F for HMG and M2F and F2M for HTG.

\subsection{Comparison systems} \label{experiment2}
\noindent\textbf{Ground truth (GT)}: The original speech audio, which serves as the upperbound.\\
\noindent\textbf{FVMVC}: Face-based memory-based zero-shot Face Voice Conversion model \cite{sheng2023face} which recently demonstrated state-of-the-art performance on LRS3 dataset.\\
\noindent\textbf{HYFace}: Our proposed method, detailed in Section \ref{sec:method}.

\subsection{Metrics} \label{experiment3}
Following Sheng \etal~\cite{sheng2023face} and other conventional VC studies, for objective evaluation, we assess the homogeneity, diversity, and objective consistency. For subjective evaluation, we examine subjective consistency, naturalness, and ABX tests. Furthermore, we propose a new evaluation metric: pitch deviation. For all objective evaluations, we randomly selected 10 source speakers for each of the 50 target speakers and repeated this process for 10 trials. This resulted in a total of 5,000 conversion pairs for each of the four evaluation sets. To measure cosine similarity, we utilized speaker embeddings generated by \textit{Resemblyzer}\footnote{https://github.com/resemble-ai/Resemblyzer}. For subjective evaluations, we use Mean Opinion Scores (MOS) collected via Amzon Mechanical Turk (MTurk). We described detailed procedure of MTurk on Supplementary B. The explanation of each metric is as follows.

\noindent\textbf{Homogeneity}: It measures cosine similarity of speaker embeddings in synthesized audio generated from different facial images of the same speaker.
Similarity value is expected to be high regardless of different view of face image on same speaker. We randomly select 10 face images from each target speaker.\\
\noindent\textbf{Diversity}: It measures cosine similarity of speaker embeddings in synthesized audio generated from different speakers. In contrast from homogeneity, here the model aims to capture distinct speaker information for different target speakers.\\
\noindent\textbf{Consistency(obj)}: It compares the speaker embedding similarity of the synthesized audio with that of the ground-truth audio from the same speaker. To ensure a robust comparison, we also assess this metric with ground-truth audio from a random speaker, referred to as `Consistency(rnd)'.\\
\noindent\textbf{Consistency(sub)}: This metric measures consistency for subjective evaluation using a 5-point MOS scale (completely inconsistent to completely consistent). It assesses whether the synthesized audio aligns with the corresponding facial images.\\
\noindent\textbf{Naturalness}: It assesses the sound quality of the synthesized audio using 5-point MOS scale (completely unnatural to completely natural).\\
\noindent\textbf{ABX test}: This evaluates the subjective preference between two models. Participants are shown a face image and asked to decide which of two synthesized audio samples, one from HYFace and the other from FVMVC, more closely matches the face in the image.\\
\noindent\textbf{Pitch deviations}: It is our newly proposed metric. Since the $F0$ is one of the key component of voice, we assess the deviation between the $F0$ of the synthesized audio and the average $F0$ (represented as $f_{0,\mathit{gt}}^{\mathit{avg}}$ in Section~\ref{sec:method}) of the ground-truth target speaker.
Note that the standard deviations (stdv) of the $F0$ for all audio samples are 29.18 Hz for male speakers and 37.50 Hz for female speakers. These values serve as a baseline, reflecting the deviation is based solely on gender class. If the model captures the associations between facial features and voice characteristics within a gender-controlled set, then it should demonstrate a deviation lower than these baseline values.

\subsection{Results}
The evaluation results for the metrics discussed in previous section can be found in Table \ref{table:result}.
As mentioned in Section \ref{experiment1}, we have created four evaluations sets: HMG (M2M and F2F), HTG (M2F and F2M). However, due to space limitations, we present the averaged scores for both HMG and HTG. For detailed results from all four evaluation sets, please refer to Supplementary C. Note that GT refers to the ground-truth audio, in which pairing between source and target of different genders (heterogeneous gender pairing) is not possible. Therefore, only HMG scores are applicable. We have not reported Consistency(obj) scores for GT as it is conceptually identical to Homogeneity.
\subsubsection{Objective results}
In Homogeneity, our proposed model HYFace presents high scores in both HMG and HTG ($p<0.01$ in paired t-tests) than FVMVC. For Diversity, the FVMVC shows better scores both in HMG and HTG ($p<0.01$ in paired t-tests). For Consistency(obj), HYFace scored higher, indicating closer similarity with the ground-truth audio.
For fair comparing, we measured the Consistency(rnd), which measures similarity between synthesized audio and the ground-truth audio of `different' speakers.
For HYFace, the Consistency(rnd) scores are 0.5577 for HMG and 0.5524 for HTG. The Consistency(obj) scores of HYFace are statistically higher than Consistency(rnd) ($p<0.01$ in paired t-tests) both in HMG and HTG, suggesting HYFace has meaningful correlation with voice characteristics of ground-truth speaker. However, in case of FVMVC, the Consistency(rnd) scores are 0.5074 for HMG and 0.5064 for HTG, showing no significant difference from its Consistency(obj) scores ($p>0.02$ for HMG and $p>0.1$ for HTG in paired t-tests). It means that there is no correlation with speaker embedding of synthesized audio from FVMVC and that of ground-truth audio.
The objective results suggest that although HYFace may show slightly poorer Diversity compared to the benchmark, its speaker embeddings significantly align with those of the ground-truth speaker, a feature not observed in the benchmark model. Additionally, our model demonstrates higher Homogeneity scores.
\subsubsection{Subjective results}
In all subjective evaluations, including Consistency(sub), Naturalness, and ABX test, our proposed HYFace model outperformed the benchmark for both HMG and HTG ($p<0.01$ in paired t-tests). Remarkably, in the Consistency(sub) metric, which assesses how well the synthesized audio matches the corresponding ground-truth facial image, HYFace achieved scores comparable to those of the ground-truth audio. Furthermore, HYFace's Naturalness score nearly approached that of the ground-truth audio. 
In terms of performance differences between HMG and HTG sets, only the Naturalness scores in the FVMVC model showed a significant decrease in the HTG set compared to HMG ($p<0.01$ in paired t-tests).

\begin{table}[t!]
\caption{Comparison of pitch deviation (in Hz) between synthesized audio and ground-truth average $F0$}
\centering
\begin{adjustbox}{width=0.36\textwidth}
\begin{tabular}{ccccc}
\hline\hline
\multirow{2}{*}{} & \multicolumn{2}{c}{HMG$\downarrow$} & \multicolumn{2}{c}{HTG$\downarrow$}\\
\cmidrule(lr){2-3} \cmidrule(lr){4-5}
 & M2M & F2F & M2F & F2M\\
\hline
FVMVC\cite{sheng2023face} & 29.55 & 34.15 & 34.75 & 28.50\\ 
HYFace & \textbf{24.01} & \textbf{29.58} & \textbf{29.15} & \textbf{24.31}\\ 
\hline\hline
\end{tabular}
\end{adjustbox}
\label{table:result2}
\vspace{-2mm}
\end{table}

\subsubsection{Pitch deviations}
Table \ref{table:result2} presents the $F0$ deviation of two models, our proposed HYFace and FVMVC, the benchmark. Across evaluation sets of all gender pairings of source and target speakers (M2M, F2F, M2F, and F2M), the proposed HYFace exhibited superior $F0$ estimation performance compared to the benchmark ($p<0.01$ in paired t-tests).
Furthermore, HYFace consistently demonstrated significantly lower deviations compared to the stdv of the ground truth. In the male cases, the deviations are 24.01 for M2M and 24.31 for F2M, both below the GT male stdv of 29.18. In female cases, the deviations are 29.58 for F2F and 29.15 for M2F, each below the GT female stdv of 37.50. It suggests that our model can nearly estimate the pitch of the target speaker based solely on facial images, even under the controlled gender set.
To our knowledge, this marks the first instance of evaluating explicit voice characteristics, pitch, associating with facial features within the face-based voice conversion domain, and also yielding meaningful results.

\section{Conclusion}
In this work, we present a novel framework for face-based voice conversion, particularly utilizing fundamental frequency estimation module, which operates solely on facial images. Through comprehensive objective and subjective evaluations, our model has achieved state-of-the-art performance. Moreover, in our newly proposed metric, which explicitly assesses the association between facial features and voice characteristics, our method has yielded meaningful results.
We hope that this research will serve as stepping stones towards providing individuals without a voice with one that fits their identity.

\section{Acknowledgements}
We sincerely give thanks to \textit{Zheng-Yan, Sheng} who is the author of \cite{sheng2023face} for dedication to the academy area and kind communication regarding model reproduction.
This work was supported by Institute of Information communications Technology Planning \& Evaluation (IITP) grant funded by the Korean government(MSIT) [No. RS-2022-II220641, XVoice: Multi-Modal Voice Meta Learning].

\bibliographystyle{IEEEtran}
\bibliography{mybib}

\end{document}

\appendix

\section{Losses}
Our proposed HYFace network is a voice conversion (VC) framework fundamentally based on conditional variational autoencoder architecture. It incorporates pretrained SSL representations as content input for the prior encoder.
To clarify our HYFace training procedure, we describe  our loss functions, which include reconstruction loss ($\mathcal{L}_{\mathit{recon}}$), KL (Kullback-Leibler) divergence loss ($\mathcal{L}_{\mathit{KL}}$), adversarial loss ($\mathcal{L}_{\mathit{adv}}(G)$, $\mathcal{L}_{\mathit{adv}}(D)$), feature matching loss ($\mathcal{L}_{\mathit{fm}}$), and two types of $F0$ loss ($\mathcal{L}_{\mathit{ff}}$, $\mathcal{L}_{\mathit{af}}$).

We use mel-spectrogram ($x_{\mathit{mel}}$) instead of waveform~($y$) for input of posterior encoder. Then, the L1 loss is used for reconstruction as follows:
\begin{equation}
\mathcal{L}_{\mathit{recon}}=||x_{\mathit{mel}} - \hat{x}_{\mathit{mel}}||_1,
\end{equation}
where $\hat{x}_{\mathit{mel}}$ refers to mel-spectrogram of $\hat{y}$, the decoder output. The prior encoder ($p_{\theta}$) takes content encoder output ($c$), which is SSL based contents embedding on our framework, and then conditioned by (face-based) speaker embedding ($s$) and frame-wise $F0$ values $f_0^i$ ($i=1,...,n$. $n$ is a number of frames) to enrich its prior distribution $z$. Note that we used linear spectrogram ($x_{\mathit{lin}}$) and speaker embedding condition $s$ for training posterior encoder ($q_{\phi}$). Then the KL divergence loss is:
\begin{equation}
\mathcal{L}_{\mathit{KL}}=\log{q_{\phi}(z|x_{\mathit{lin}},s)-\log{p_{\theta}(z|c,f,s)}}.\\
\end{equation}
We designed a loss for $f_0^i$ using speaker-wise average $F0$ ($f_{0,\mathit{gt}}^{\mathit{avg}}$), $c$, and $s$. The loss $\mathcal{L}_{\mathit{ff}}$ is as follows and $\mathit{FF}$ denotes Frame-wise $F0$ Decoder.
\begin{equation}
\mathcal{L}_{\mathit{ff}}=\frac{1}{n}\sum_{i=1}^n(f_{0,\mathit{gt}}^i-\mathit{FF}(f_{0,\mathit{gt}}^{\mathit{avg}}, c, s))^2,
\end{equation}
where $f_{0,\mathit{gt}}^i$ refers to the ground truth frame-wise $F0$ values. Note that $f_{0,\mathit{gt}}^{\mathit{avg}}$ value represents the average of $f_{0,\mathit{gt}}^i$ values across all audio frames for each speaker in the training dataset.

For stable quality for our generated waveform $\hat{y}$ which is output of the decoder, we used adversarial loss ($\mathcal{L}_{\mathit{adv}}(G)$ and $\mathcal{L}_{\mathit{adv}}(D)$) and feature matching loss ($\mathcal{L}_{\mathit{fm}}$):
\begin{subequations}
\begin{align}
\mathcal{L}_{\mathit{adv}}(D)&=\mathbb{E}_{(y,\hat{y})}[(D(y)-1)^2+(D(\hat{y}))^2],\\
\mathcal{L}_{\mathit{adv}}(G)&=\mathbb{E}_{(\hat{y})}[(D(\hat{y})-1)^2],\\
\mathcal{L}_{\mathit{fm}}&=\mathbb{E}_{(y,\hat{y})}[\sum_{i=1}^T\frac{1}{N_l}||D^l(y)-D^l(\hat{y})||_1],
\end{align}
\end{subequations}
where $D$, $G$, $D^l$ denotes the discriminator, generator and $l$-th layer of the discriminator, respectively. $T$ and $N_l$ denotes the total number of layers in $D$ and number of features, respectively.

Finally, total loss $\mathcal{L}_{\mathit{vc}}$ of our face-based voice conversion network is as follows:
\begin{equation}
\mathcal{L}_{vc}=\mathcal{L}_{\mathit{recon}}+\mathcal{L}_{\mathit{KL}}+\mathcal{L}_{\mathit{adv}}(G)+\mathcal{L}_{\mathit{fm}}+\mathcal{L}_{\mathit{ff}}.
\end{equation}

Also, we train a face-based average $F0$ estimation network ($\mathit{AF}$) independently with face-based voice conversion network to predict $f_{0,\mathit{gt}}^{\mathit{avg}}$ from face image $v$. The loss $\mathcal{L}_{\mathit{af}}$ is as follows:
\begin{equation}
\mathcal{L}_{\mathit{af}}=(f_{0,\mathit{gt}}^{\mathit{avg}}-\mathit{AF}(v))^2.
\end{equation}

\begin{table}[h]
\caption{Evaluation result for Homogeneity}
\centering
\begin{adjustbox}{width=0.4\textwidth}
\begin{tabular}{ccccc}
\hline\hline
\textbf{Homogeneity}$\uparrow$ & M2M & F2F & M2F & F2M\\
\hline
GT & 0.7412 & 0.7499 & - & - \\ \hline
FVMVC & 0.6404 & 0.6379 & 0.6372 & 0.6430 \\ 
HYFace & \textbf{0.6860} & \textbf{0.6680} & \textbf{0.6719} & \textbf{0.6867} \\ 
\hline
\hline
\end{tabular}
\end{adjustbox}
\label{table:result1}
\end{table}
\vspace{-4mm}

\begin{table}[h]
\caption{Evaluation result for Diversity}
\centering
\begin{adjustbox}{width=0.4\textwidth}
\begin{tabular}{ccccc}
\hline\hline
\textbf{Diversity}$\downarrow$ & M2M & F2F & M2F & F2M\\
\hline
GT & 0.5418 & 0.5423 & - & - \\ \hline
FVMVC & \textbf{0.5982} & \textbf{0.5903} & \textbf{0.5930} & \textbf{0.6021} \\ 
HYFace & 0.6122 & 0.6022 & 0.6084 & 0.6122 \\ 
\hline
\hline
\end{tabular}
\end{adjustbox}
\label{table:result2}
\end{table}

% \newpage
\section{Crowdsource Evaluation}
\begin{figure*}[htp] 
    \centering
    \subfigure{\includegraphics[width=0.95\linewidth]{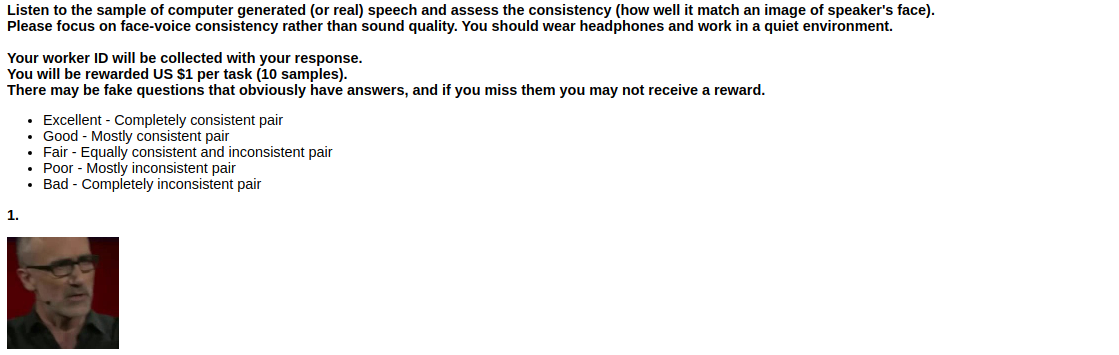}}
    \subfigure{\includegraphics[width=0.95\linewidth]{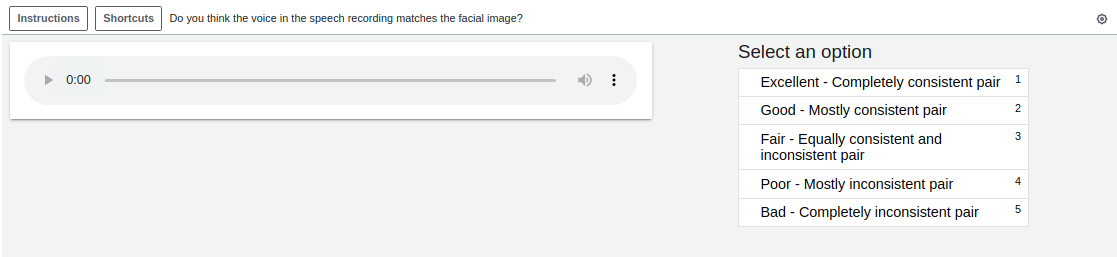}}
\caption{Instruction of Consistency(sub)}
\label{fig:consistency}
\end{figure*}

\begin{figure*}[htp]
	\centering
	\includegraphics[width=0.95\textwidth]{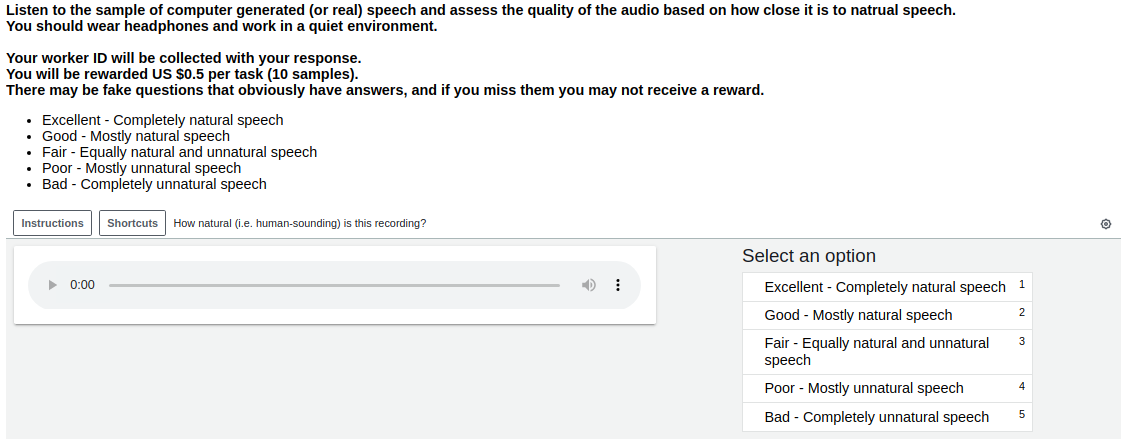}
        \caption{Instruction of Naturalness}
	\label{fig:naturalness} 
\end{figure*}

\begin{figure*}[p]
\centering
    \subfigure{\includegraphics[width=0.95\linewidth]{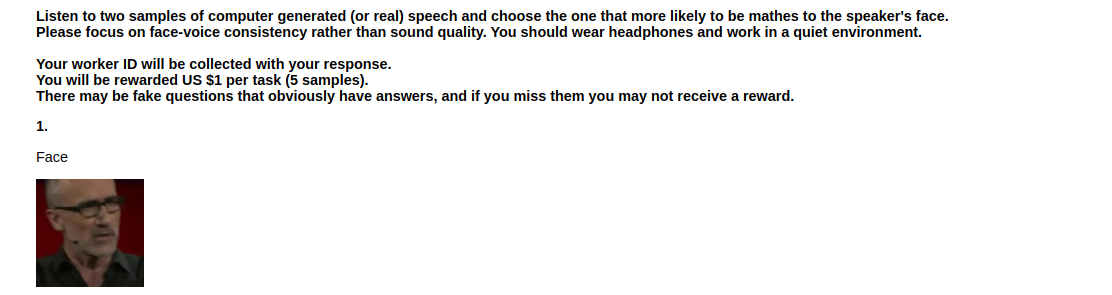}}
    \subfigure{\includegraphics[width=0.95\linewidth]{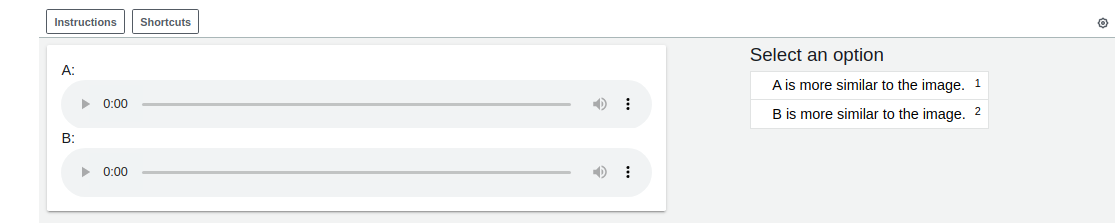}}
    \caption{Instruction of ABX test}
    \label{fig:abx}
\vspace{5in}
\end{figure*}

For subjective evaluations, we use Mean Opinion Scores (MOS) collected via Amzon Mechanical Turk (MTurk). The metrics are subjective consistency, naturalness, and ABX test.
Here we attach the webpage instructions for 3 crowdsource evaluations based on MTurk. The instructions for subjective consistency, naturalness, and ABX test are shown in Figure \ref{fig:consistency}, Figure \ref{fig:naturalness}, and Figure \ref{fig:abx}, respectively. The task unit of MTurk is called Human Intelligence Task (HIT).\\

\noindent\textbf{Consistency(sub)}: This metric measures consistency for subjective evaluation using a 5-point MOS scale (completely inconsistent to completely consistent). It assesses whether the synthesized audio aligns with the corresponding facial images.
For each evaluation, `HIT' consists of 10 audio clips paired with their corresponding facial images from the same speaker. These include: one ground-truth audio, one fake audio, four synthesized audio from our proposed model HYFace, and four from the benchmark model. Four synthesized audio clips from each conversion model is randomly selected on each of all possible gender pairings (male to male, male to female, female to male, female to female). The inclusion of a fake audio clip serves to identify unreliable respondents. Each HIT is distributed to 100 subjects and the reward for each HIT was 1 USD. The total amount of budget we spent for subjective consistency was therefore, 100 USD. The webpage instruction on MTurk is in Figure~\ref{fig:consistency}.\\

\noindent\textbf{Naturalness}: It assesses the sound quality of the synthesized audio using 5-point MOS scale (completely unnatural to completely natural). Each `HIT' consists of 10 audio clips (no facial images). These include: one ground-truth audio, one fake audio, four synthesized audio from our proposed model HYFace, and four from the benchmark model. Then each HIT is distributed to 100 subjects and the reward for each HIT was 0.5 USD. The total amount of budget we spent for subjective consistency was therefore, 50 USD. The webpage instruction on MTurk is in Figure~\ref{fig:naturalness}.\\

\begin{table*}[t]
\caption{Evaluation result for Consistency(obj), similarity with the ground-truth audio from the ground-truth speaker. The Consistency(rnd) scores, indicating the similarity to the ground-truth audio from a random speaker, are shown in parentheses. `**' indiciates that Consistency(obj) has statiscally significant high value than Consistency(rnd) values ($P<0.01$ in paired $t$-test)}
\centering
% \begin{adjustbox}{width=0.9\textwidth}
\begin{tabular}{ccccc}
\hline\hline
\textbf{Consistency(obj)}$\uparrow$ & M2M & F2F & M2F & F2M\\
\hline
FVMVC & 0.5096 (0.5105) & **0.5115 (0.5043) & **0.5077 (0.5014) & 0.5094 (0.5114) \\ 
HYFace & **\textbf{0.5695} (0.5565) & **\textbf{0.5697} (0.5589) & **\textbf{0.5633} (0.5528) & **\textbf{0.5632} (0.5524) \\ 
\hline
\hline
\\
\end{tabular}
% \end{adjustbox}
\label{table:result3}
\end{table*}

\noindent\textbf{ABX test}: This evaluates the subjective preference between two models. Participants are shown a face image and asked to decide which of two synthesized audio samples, one from HYFace, the proposed method, and the other from FVMVC, the benchmark, more closely matches the face in the image.
Each `HIT' in the ABX test comprises 5 sets. Each set consists of one facial image paired with two synthesized audio clips: one from HYFace and one from FVMVC.
Then each HIT is distributed to 100 subjects and the reward for each HIT was 1 USD.
The total amount of budget we spent for ABX test was therefore, 100 USD. The webpage instruction on MTurk is in Figure~\ref{fig:abx}.\\

\section{Detailed results}
This parts present the objective and subject results on four evaluation sets. M2M for male-to-male conversion, F2F for female-to-female conversion, M2F for male-to-female conversion, and F2M for female-to-male conversion.

\subsection{Objective results}
In Homogeneity, our proposed model HYFace presents high scores in all four evaluation sets (M2M, F2F, M2F, and F2M) ($p<0.01$ in paired t-tests) than FVMVC. The results is on Table~\ref{table:result1}. For Diversity, the FVMVC shows better scores in all four datasets ($p<0.01$ in paired t-tests) than HYFace and the results is on Table~\ref{table:result2}.

The result of objective consistency (Consistency(obj)) which indicates the speaker embedding similarity between the ground-truth audio and the synthesized audio from same speaker is in Table~\ref{table:result3}, HYFace scored higher than the benchmark. It indicates that HYFace has closer similarity with audio from the ground-truth speaker. Also, for all four datasets M2M, F2F, M2F, and F2M, HYFace showed statistically higher than Consistency(rnd) which indicates the speaker embedding similarity between the synthesized audio and the ground-truth audio from random speaker ($p<0.01$ in paired t-tests), confirming HYFace has meaningful correlation with voice characteristics of ground-truth speaker again. FVMVC showed statistically higher than Consistency(rnd) on two of four datasets, F2F and M2F datasets, the model may have some bias to generating speaker embedding related to the characteristics of female gender.

\begin{table}[h]
\caption{Evaluation result for Consistency(sub)}
\centering
\begin{adjustbox}{width=0.4\textwidth}
\begin{tabular}{ccccc}
\hline\hline
\textbf{Consistency(sub)}$\uparrow$ & M2M & F2F & M2F & F2M\\
\hline
GT & 3.9348 & 4.1591 & - & - \\ \hline
FVMVC & 3.5778 & 3.5632 & 3.4615 & 3.5402 \\ 
HYFace & \textbf{3.9444} & \textbf{3.8387} & \textbf{3.8202} & \textbf{3.8172} \\ 
\hline
\hline
\end{tabular}
\end{adjustbox}
\label{table:result4}
\end{table}
\vspace{-2mm}

\begin{table}[h]
\caption{Evaluation result for Naturalness}
\centering
\begin{adjustbox}{width=0.4\textwidth}
\begin{tabular}{ccccc}
\hline\hline
\textbf{Naturalness}$\uparrow$ & M2M & F2F & M2F & F2M\\
\hline
GT & 4.000 & 3.8095 & - & - \\ \hline
FVMVC & 3.4337 & 3.3855 & 3.2651 & 3.2289 \\ 
HYFace & \textbf{3.8434} & \textbf{3.8193} & \textbf{3.7108} & \textbf{3.8193} \\ 
\hline
\hline
\end{tabular}
\end{adjustbox}
\label{table:result5}
\end{table}

\begin{table}[h]
\caption{Evaluation result for ABX test(\%)}
\centering
% \begin{adjustbox}{width=0.4\textwidth}
\begin{tabular}{ccccc}
\hline\hline
\textbf{ABX test}$\uparrow$ & M2M & F2F & M2F & F2M\\
\hline
FVMVC & 0.37 & 0.42 & 0.41 & 0.43 \\ 
HYFace & \textbf{0.63} & \textbf{0.58} & \textbf{0.59} & \textbf{0.57} \\ 
\hline
\hline
\end{tabular}
% \end{adjustbox}
\label{table:result6}
\end{table}

\begin{table}[h]
\caption{Evaluation results for pitch deviation (in Hz)}
\centering
% \begin{adjustbox}{width=0.36\textwidth}
\begin{tabular}{ccccc}
\hline\hline
\textbf{Pitch dv}$\downarrow$ & M2M & F2F & M2F & F2M\\
\hline
FVMVC & 29.55 & 34.15 & 34.75 & 28.50\\ 
HYFace & \textbf{24.01} & \textbf{29.58} & \textbf{29.15} & \textbf{24.31}\\ 
\hline\hline
\end{tabular}
% \end{adjustbox}
\label{table:result7}
\vspace{-2mm}
\end{table}

\subsection{Subjective results}
In all subjective evaluations, including Consistency(sub) (Table~\ref{table:result4}), Naturalness (Table~\ref{table:result5}), and ABX~ test (Table~\ref{table:result6}), our proposed HYFace model outperformed the benchmark ($p<0.01$ in paired t-tests). Remarkably, in the Consistency(sub) metric, which assesses how well the synthesized audio matches the corresponding ground-truth facial image, HYFace achieved scores higher than the ground truth on M2M dataset. Furthermore, HYFace's Naturalness score nearly approached that of the ground-truth audio and achieved higher scroes than the ground truth on F2F dataset.

\subsection{Pitch deviations}
Table \ref{table:result7} presents the $F0$ deviation of two models, HYFace and FVMVC. Across datasets of all gender pairings of source and target speakers (M2M, F2F, M2F, and F2M), the proposed HYFace exhibited superior $F0$ estimation performance compared to the benchmark ($p<0.01$ in paired t-tests).
For the detailed discussion, please refers to our main paper.

\subsection{Comparative performance across the evaluation sets}
Our study encompasses four datasets, M2M, F2F, M2F and F2M, covering all possible male and female pairings for source and target speakers. Specifically, we hypothesized that the M2F and F2M evaluations sets, which involve heterogeneous gender pairings between source and target speakers, will exhibit reduced performance compared to the homogeneous M2M and F2F pairings.
The results for Consistency(obj) and Consistency(sub) both indicate a drop in performance when the source gender changes to a heterogeneous gender from the target's (from M2M to F2M and from F2F to M2F). The results in Naturalness also show similar decline. Both results support our hypothesis to some extent.
Interestingly, we observed some performance discrepancies based on the gender of the target speaker, regardless of the gender of source speaker. We plan to analyze further in future work.

\newpage

% \bibliographystyle{IEEEtran}
% \bibliography{mybib}